\documentclass[12pt,pdftex]{article} \usepackage{rotating}
\usepackage{natbib} \usepackage{graphicx} 
\usepackage{amsmath} 

 \title{Predicting trophic relations in
  ecological networks: a test of the Allometric Diet Breadth Model}
\author{Stefano Allesina\\ \small
  Dept. of Ecology \& Evolution,\\
  \small Computation Institute,\\
  \small The University of Chicago\\
  \small \texttt{sallesina@uchicago.edu}
  \normalsize
}
 \bibpunct{(}{)}{,}{a}{}{;}
\begin{document}
\maketitle
\baselineskip 24pt
\begin{abstract}
  Few of food web theory hypotheses/predictions can be readily tested
  using empirical data. An exception is represented by simple
  probabilistic models for food web structure, for which the likelihood
  has been derived. Here I test the performance of a more complex model
  for food web structure that is grounded in the allometric scaling of
  interactions with body size and the theory of optimal foraging
  (Allometric Diet Breadth Model - ADBM). This deterministic model has
  been evaluated measuring the fraction of trophic relations correctly
  predicted. I contrast this value with that produced by simpler models
  based on body sizes and find that the data does not favor the more
  complex model: the information on allometric scaling and optimal
  foraging does not significantly increase the fit to the data. Also, I
  take a different approach and compute the p-value for the fraction of
  trophic interactions correctly predicted by ADBM with respect to three
  probabilistic null models. I find that the ADBM is clearly better at
  predicting links than random graphs, but other models can do even
  better. Although optimal foraging and allometric scaling could improve
  our understanding of food webs, the models need to be ameliorated to
  find support in the data.
\end{abstract}

\section*{Introduction}
Understanding the main forces shaping the topology of food webs
(networks depicting who eats whom in an ecosystem) is a central problem
in ecology that has received a lot of attention
\citep{Cohen90FW,WilliamsM00,Cattin2004,Allesina2008Science,AllesinaGroups}.
This problem has been typically investigated using simple probabilistic
models, but recently models that incorporate explicitly relevant
biological quantities in their assumptions have started appearing in the
literature \citep{Loeuille2005,Rossberg2006,Petchey2008}.

In a work that investigated the role of body size and optimal foraging
theory in shaping food web structure, \cite{Petchey2008} assessed the
goodness of variants of their main model, measuring the proportion of
empirical connections a model is able to predict. If a model proposes
$K$ connections among species of which $M$ are present in the empirical
data set, then the proportion of correct links (overlap) is
$\Omega=M/K$. They measured this overlap for their Allometric Diet
Breadth Model (ADBM), and they showed that the best version of the ADBM
is able to correctly predict, depending on the empirical network
examined, between $5\%$ and $65\%$ of the proposed links
\citep{Petchey2008}. The ADBM is based on two main ideas: optimal
foraging theory and allometric scaling of relevant quantities with body
size \citep{Beckerman2006,Petchey2008}. The ADBM is different from most
previous models also because it is not probabilistic: given an empirical
network, body sizes for all the species and number of links in the
network, 4 further parameters dealing with the foraging are optimized
numerically and a single network is produced deterministically. Here I
compare the ADBM with simpler deterministic models that include
information on body size but do not make use of allometric scaling and
optimal foraging, and I find that the data does not support the use of
the more complex model.

Also, I derive a p-value for the $\Omega$ produced by the ADBM using as
a reference a random digraph \citep{Erdos1960}, a variation of the
cascade model \citep{Cohen90FW} and a recently proposed group-based
random digraph \citep{AllesinaGroups}. The derivation of the probability
mass function for these simple models is a step forward in the analysis
of more complex models for food web structure, for which the derivation
of a likelihood can be almost impossible. The derivation presented here
can help associating statistical significance to the results of highly
complex models, such as those based on evolving networks or systems of
differential equations \citep{McKane98}.

Results show that the ADBM performs significantly better than the random
graph in terms of overlap. It also performs better than the cascade
model analyzed here in most of the cases. The performance is
significantly worse than that of the group-based random digraph. 

In summary, even though allometric scaling and optimal foraging have the
potential to illuminate the topology of food webs, the present models do
not provide enough evidence to support this claim.

\section*{Methods}
\subsection*{The Allometric Diet Breadth Model}
Here I briefly describe the ADBM in its ``Ratio'' incarnation, that is
the one that produces the best fit to the empirical data. A more
detailed description of the model and its variations can be found in the
original articles \citep{Beckerman2006,Petchey2008}.

The model takes as an input a vector $\vec{B}$ describes the species
body sizes. The model requires the number of links ($L$) for the
empirical food web one wants to replicate that will be used in the
numerical optimization routine. Then, the model uses four other
parameters $a$, $a_1$, $a_2$ and $b$ that determine the foraging
behavior of the species in the food web. The model consists of two
steps: a) compute, for each predator, the profitability of each possible
prey; b) compute a diet breadth (i.e. number of prey) for each predator:
this number is chosen to maximize the rate of energy intake.  Repeating
the two steps for all the consumers produces a food web.

Here is a detailed description of the two steps outlined above:
\begin{enumerate}
\item Profitability. The profitability $P_{ij}$ of prey $i$ for consumer
  $j$ is defined as:
  \begin{equation}
    P_{ij}=\frac{B_i(b B_j -B_i)}{B_j}
  \end{equation}
  where $B_i$ is the body size of species $i$ and $b$ is a positive
  parameter.
\item Diet Breadth. A predator $j$ will prey upon $z$ species, where $z$
  is the value in $0, 1, \ldots, S$ that maximizes the function:
  \begin{equation}
    f(z,j)=\frac{a \sum_{i={\sigma_1}}^{\sigma_z} B_i^{\left(\frac{1}{4}+a_1\right)}B_j^{a_2}}
    {1+a \sum_{i=\sigma_1}^{\sigma_z}\frac{B_i^{\left(-\frac{3}{4}+a_1\right)}B_j^{a_2}}{b-\frac{B_i}{B_j}}}
  \end{equation}
  where the permutation $\sigma$ is the permutation that orders the prey
  according to decreasing profitability: if $z=1$ is the value that
  maximizes $f(z,j)$, then consumer $j$ will choose only the most
  profitable prey, if $z=2$ then it will choose the two most profitable
  prey and so on. This apparently complicated function is easily
  justifiable in terms of optimal foraging. The three parameters $a$,
  $a_1$ and $a_2$ are needed for computing the attack rate, and the
  parameter $b$ is involved in the computation of the handling time.
\end{enumerate}
Repeating the two steps for all consumers generates a food web that will
be compared with the empirical data.  The performance of the model is
measured as the fraction of links that correctly match the ones in the
empirical food web. If an instance of the ADBM for a given network
produces $K$ links of which $M$ are present in the empirical food web,
then the proportion of links correctly predicted, or overlap is
$\Omega=M/K$. The parameters $a$, $a_1$, $a_2$ and $b$ are optimized
numerically so that the model a) correctly predicts the total number of
links in the network ($K \approx L$) and b) $\Omega$ is maximized.

Running the ADBM for the 9 published food webs examined here yields
$\Omega \in [0.08,0.65]$ (Table 1).

\subsection*{Four simple models based on body size}
One of the characteristics of the ADBM is that it produces interval
networks: when species are ordered according to body size all the prey
of a given predator are adjacent. Food webs are known to be
quasi-interval
\citep{WilliamsM00,Cattin2004,StoufferInterval,Allesina2008Science}, and
this could be a main driver of the performance of the ADBM. It makes
sense therefore to compare its performance with that of models that
retain the intervality but do not contain extra information regarding
optimal foraging and allometric scaling. Of all possible models, I
analyze here four that have the virtue of being very simple and sharing
the same structure. For each possible predator-prey couple, one computes
a value that depends on body sizes of predator and prey: $z_{ij}={\it
f}(B_i,B_j)$. If $a \leq z_{ij}< b$, where $a$ and $b$ are food
web-dependent parameter estimates, one draws a connection. If $z_{ij}$
is not included in the interval $(a,b]$, no connection is drawn. In what
follows, I analyze four different ${\it f}(B_i,B_j)$:
\begin{enumerate}
\item ``Diff'': ${\it f}(B_i,B_j)=B_i-B_j$. The difference between
  predator ($i$) and prey ($j$) sizes must fall in $(a,b]$ to draw a
  connection.
\item ``Ratio'': ${\it f}(B_i,B_j)=B_i/B_j$. The ratio between the body
  sizes is what drives the structure of the food web.
\item ``LogRatio'': ${\it f}(B_i,B_j)=ln(B_i+1)/ln(B_j+1)$. Where 1 is
  added so that the function is positive for all possible body sizes.
\item ``DiffRatio'': ${\it f}(B_i,B_j)=(Bj-Bi)(B_i/B_j)$. This model
  combines the first two models. Note that the function is very similar
  to Eq. 1 (profitability).
\end{enumerate}
All the four models produce interval networks, are deterministic in
nature (as the ADBM), and require the optimization of two parameters
$(a,b)$ that can be easily accomplished by trying all relevant
combinations.

For each model/food web, I optimize $a,b$ so that a) the number of links
produced is similar to the one measured empirically: if the ADBM
proposes $K$ links and $|K-L|=t$, I accept as possible solutions only
those whose number of connections is in $[L-t,L+t]$ (this is to ensure
that the comparison is fair). b) Among all solutions satisfying the
previous requirement, I choose the one that maximizes
$\Omega$. Contrasting these values with those produced by the ADBM can
help us determine whether optimal foraging and allometric scaling do
play a crucial role in predicting the links in the food web.

\subsection*{p-value: a random digraph}
Another way to assess the goodness of a given $\Omega$ is to associate a
p-value to it. This quantity expresses the probability of obtaining a
result that is equally good or better using a null model. In the
remainder of the section I derive analytically such a p-value when the
null model is a random digraph, while in the Appendix I derive the
p-value when the null model is a cascade model or a group-based random
digraph. I chose these models because they share the same derivation,
and are in a continuum of complexity that makes the comparison easier.

A random digraph \citep{Erdos1960} is the simplest possible way to
produce networks: it takes just two parameters ($S$, the number of nodes
in the network - standing for species, and $p$, the probability that two
species are connected by feeding relations) and produces a network
connecting any two species with a directed link with probability $p$. We
want to know the probability ${\cal P}(M,K|S,p,N(S,L))$ that a random
graph using parameters $S$ and $p$ produces a network with $K$ links, of
which $M$ are matching those of an empirical network $N$ that contains
$S$ species and $L$ links. We can start by writing the probability that
the random graph produces exactly $K$ links. This is a binomial
probability mass function (pmf):

\begin{equation}
{\cal P}(K|S,p,N(S,L))=\binom{S}{K} p^{K}(1-p)^{(S^2-K)}
\end{equation}

If we set $p=L/S^2$ we maximize the probability of obtaining $L$ links
in the generated network (this is also the maximum likelihood estimate
for the parameter). Once we know that the graph has produced $K$ links,
we can compute the probability that of these $M$ are matching those of
the empirical network $N(S,L)$ using a hypergeometric distribution:

\begin{equation}
{\cal P}(M|S,p,N(S,L),K)=\frac{\binom{L}{M}\binom{S^2-L}{K-M}}{\binom{S}{K}}
\end{equation}

The joint bivariate pmf becomes:

\begin{equation}
  {\cal P}(M,K|S,p,N(S,L))=p^{K}(1-p)^{(S^2-K)}\binom{L}{M}\binom{S^2-L}{K-M}
  \label{LikelihoodRnd}
\end{equation}

This pmf assumes values for $K \in [0,\ldots,S^2]$ and, for each $K$, $M
\in [0,\ldots, min(K,L)]$. We can therefore describe the pmf in a table
with $(S^2+1)(L+1)-L(L+1)/2$ values associated with all the possible
combinations of $K$ and $M$. An example of such a table is reported in
Figure 1 for a small network. The table expressing the bivariate
probability mass function shows the probability of obtaining any
combination of $K$ and $M$. Because we are interested in the pmf for
$\Omega=M/K$ we can map the results from the bivariate pmf into a
univariate distribution by summing the probabilities for all the
combinations of $K$ and $M$ leading to the same $\Omega$. For example,
in Figure 1 I report the first few rows of such a table. From this, one
can draw the complete pmf for $\Omega$.

Deriving the probability of reproducing exactly the data shows the
relation between $\Omega$ and the likelihood. In fact, the likelihood
can be seen as the probability of having $\Omega=1$ when $M=K=L$. By
substituting in Eq. \ref{LikelihoodRnd} we obtain:

\begin{equation}
  {\cal P}(L,L|S,p,N(S,L))=p^{L}(1-p)^{(S^2-L)}={\cal L}(S,p|N(S,L))
\end{equation}

We can readily write also the expression for the AIC, whose values will
be used in the Discussion. The number of parameters of the model is
$\theta=2$. The Akaike's Information Criterion \citep{AIC} becomes:

\begin{equation}
  AIC=2\theta-2log{\cal L}= 4-2(L\, log(p)+(S^2-L)\,log(1-p))
\end{equation}

\section*{Results}
I replicated the results obtained by Petchey et al. \citep{Petchey2008}
for nine food webs: Benguela Pelagic \citep{Yodzis98}, Broadstone Stream
\citep{Broadstone}, Scotch Broom \citep{MemmottJournalofPre2000},
Carpinteria Salt Marsh \citep{Lafferty2006}, Coachella Valley
\citep{Polis1991}, Sierra Lakes \citep{SierraLakes}, Skipwith Pond
\citep{Warren1989}, Tuesday Lake \citep{Tuesday} and Ythan Estuary
\citep{HALL91}. The optimized parameters for the ADBM were taken from
the original article \citep{Petchey2008} so that for the produced
network the number of connections match that of the corresponding
empirical network and the overlap is maximized. I then analyzed the same
networks using the four simple models based on body sizes presented
above. I am reporting in Table 1 all the overlap values. In three cases
the ADBM is the best performing model (including 2 ties). In the other
cases one or more models have higher $\Omega$ than the ADBM. Each of the
four models produces the highest $\Omega$ in three cases (including
ties). For the ``Broom'' system all the four models have better overlap
than the ADBM. The ``Diff'' model shows higher or equal $\Omega$ values
for 5 networks. The ``LogRatio'' in 4 cases. The other two models yield
higher or equal values in 3 cases.

In all cases the results are quite similar to those produced by the
ADBM, as confirmed when the exact location of predicted and
non-predicted links is examined (Figures 2 and 3): the models tend to
correctly predict the same links and fail in the same regions of the
matrix. The similarity with the ADBM is particularly pronounced for the
``Ratio'', and ``LogRatio'', while the ``Diff'' model tends to select a
different set of links compared to the other models. In no case any of
the models predicted exactly the same links.

Note however that the four simpler models optimize 2 parameters, while
the ADBM requires 4 parameters. The ADBM is therefore more flexible and
this should lead to better performance. How can we then fairly compare
the models? If these were probabilistic models, then we could use for
example AIC (or BIC, or any other selection criteria) to balance model
performance and complexity. No simple solution however exists for
deterministic models. One possibility is therefore to make the models
probabilistic. This can be done in a straightforward way. Every time a
deterministic model would draw a link, we can instead draw it with
probability $q_1$. If the deterministic model does not predict a link,
we can still draw it in the probabilistic counterpart with probability
$q_2$. Deriving the likelihood for such a process is a simple extension
of that of the models presented above, and we can see that the maximum
likelihood estimates for $q_1$ and $q_2$ are $\Omega=M/K$ and
$(L-M)/(S^2-K)$ respectively. While this modification makes all the
models general (i.e. they can produce any network), it also negatively
affect the expected $\Omega$ value. For a deterministic model $X$ that
proposes $K$ links of which $M$ are present in the empirical network,
the expected $\Omega$ for its probabilistic version $X'$ is:

\begin{equation}
  E\left[\underset{X'}{\Omega} \right] = \frac{Mq_1+(L-M)q_2}{Kq_1+(S^2-K)q_2}
  =\frac{M^2 S^2 + L^2 K - 2LMK}{LK(S^2-K)}
\end{equation}

For example, if the ADBM yields $\underset{ADBM}{\Omega}=0.57143$ for
the Benguela food web in the deterministic case, the probabilistic
version yields $E\left[\underset{ADBM'}{\Omega}\right]=0.37843$, a
decrease of 1/3 in performance. Nevertheless, this allows a fair
comparison among the models by means, for example, of AIC. The values
are reported in Table 2. When we account for model complexity, the
probabilistic version of the ADBM never yields the best AIC, the
``Diff'' has the best value in 4 cases and the remaining 5 cases are
split among the remaining models. The use of AIC allows also the use of
``Akaike weights'' \citep{burnham}. These quantities provide a measure
of strength of the evidence for each model. The results are reported in
Table 2 and show that we can say with confidence that the ADBM is not
the best among the models in all cases but three (Benguela, Skipwith and
Tuesday, $A.W. \geq 0.05$). In no case we find strong evidence for the
ADBM ($A.W. \geq 0.95$).

I also computed the probability of obtaining an $\Omega$ that is greater
or equal than that of the ADBM using the random graph, cascade model and
group based model (Methods, Appendix). In all these cases, I chose
parameters that a) made the expected number of links $E[K]=L$ and b)
minimized the AIC. Note that this optimization does not target the
overlap directly. For the random graph the optimization is simply done
by setting $p=L/S^2$. For the cascade model, I searched using a genetic
algorithm the best hierarchy that maximized the likelihood. The two
parameters were set to $p_U=2L_U/(S(S-1))$ and $p_L=2L_L/(S(S+1))$ to
maximize the likelihood and obtain on average $L$ links. The same type
of search can be performed for the group-based random graph. Also here,
I tried to find the configuration with the minimum AIC. While in the
cascade model the number of parameters is fixed (and therefore
maximizing the likelihood minimizes AIC), in this model the number of
parameters varies according to the number of groups $\gamma$. I
therefore searched, following \cite{AllesinaGroups}, for the balance
between the number of parameters and goodness of fit using Akaike's AIC
\citep{AIC}. The results in terms of likelihoods, number of parameters
and AIC values are reported in Table 3.

For each model, I computed the expected overlap with the data
($E[\Omega]$) and the probability that a model $x$ produces an overlap
value equal or greater than that of the ADBM $\left({\cal
    P}\left(\underset{x}{\Omega}\geq \underset{ADBM}{\Omega}
  \right)\right)$ (Table 1). I computed these quantities analytically
for the random graph ($RND$) and cascade ($CASC$) models. Because
listing all combinations for the group-based case ($GROUP$) is not
computationally feasible, I constructed $10^5$ networks for each data set
using this model, and I measured the overlap in this set of generated
networks.

\subsection*{Discussion}
I contrasted the ADBM with four deterministic models that retain
intervality (predators prey upon consecutive species) and information on
body sizes, but do not include optimal foraging and allometric
scaling. I found that these models perform as well as or even better
than the ADBM. This is true regardless the specific analysis performed
(i.e. $\Omega$ values, AIC of the probabilistic counterpart of each
model, Akaike weights, direct inspection of the predicted links). The
results indicate that including allometry and optimal foraging, although
biologically realistic, does not improve the fit to the data. This can
be happening either because these features do not leave a strong
signature in food web structure or because they have not been correctly
included in the models. Also, the similarity among the results of the
simpler models (especially ``Ratio'' and ``LogRatio'') and the ADBM is
so strong that one may suspect that the results of the ADBM are totally
driven by simpler mechanisms. In particular, intervality accounts for
most of the successes and failures of these simple models in predicting
links. Note however that possibly using body size is not the way of
ordering the species that maximizes intervality: if we were to find the
best species' trait that maximizes diet intervality, we could build
models such as the ones illustrated above that would yield a better fit
to the empirical data.

By examining $p-$values I found that the ADBM performs, in terms of
overlap, significantly better than the random digraph in all cases \\
$\left({\cal P}\left(\underset{RND}{\Omega}\geq \underset{ADBM}{\Omega}
\right)<<0.05\right)$. With respect to the cascade model presented in
the Appendix, the ADBM performs significantly better in 7 cases, and
yields non-significant results in two cases (Broom, ${\cal P}>0.06$ and
Skipwith ${\cal P}>0.45$). The group-based model performs significantly
better than the ADBM (${\cal P}\simeq 1.0$ in all cases). These results
are exactly reflected also in the expected values for the overlap of the
three models: the random graph on average presents much lower overlap
than the ADBM (mean difference between models $=-0.232$), the cascade is
better than the random (mean difference with the ADBM $=-0.13$) and the
group-based model does much better than the ADBM (mean difference
$=0.292$). These results are hardly surprising, given that they mirror
perfectly the complexity of the models: the random and cascade have less
parameters than the ADBM, while the group-based has many more. AIC (or
BIC, or other criteria) for probabilistic models can deal with the
assessment of the goodness of fit of a model accounting for both its
performance and its complexity: a model has to do much better in terms
of performance to justify a greater number of parameters. AIC is well
rooted in the information theory, being a measure of information loss
when the model is used instead of the data. Of the three probabilistic
models presented here, the group-based has better overlap, likelihood
and AIC in all cases (Table 4). Note that the AIC for the probabilistic
version of the ADBM presented above is worse than that of the random
case in 5 cases, and worse than the cascade in all cases. This means
that the straightforward way of making the model probabilistic greatly
hampers its performance. Producing a better model grounded in optimal
foraging theory that is probabilistic in nature is definitely possible,
and should be pursued to test whether these mechanisms could contribute
to our understanding of network structure.

The results of this exercise also show that measuring overlaps without a
quantitative comparison with other models is far from being
satisfactory. Accepting these numbers at face value without including
the probability of obtaining them using simpler models or even at random
can lead us to finding patterns and results that vanish once we
scrutinize the models in detail. In order to test whether and how
optimal foraging, allometric scaling or any other mechanism do influence
food web structure, a rigorous statistical analysis such as the one
presented here is required. Based on the data, one can conclude that in
order to prove that optimal foraging and allometric scaling are
important for food web structure, they need to be embedded in better
models than the current ones. In the meantime, for lack of a better
alternative we cannot reject the null hypothesis that these forces play
no role in shaping food webs.

\section*{Acknowledgments}
I wish to thank O.L. Petchey for providing the data necessary to
replicate the ADBM results and for interesting discussion. Two anonymous
referees provided useful comments.
Part of this work was carried out when S.A. was a postdoctoral associate
at the National Center for Ecological Analysis and Synthesis, a center
funded by National Science Foundation grant DEB-0072909, and the
University of California, Santa Barbara. This work was supported by NSF
grant EF-0827493.

\bibliography{Bib_Overlap}

\newpage

\begin{figure}
  \begin{center}
    \includegraphics[width=\textwidth]{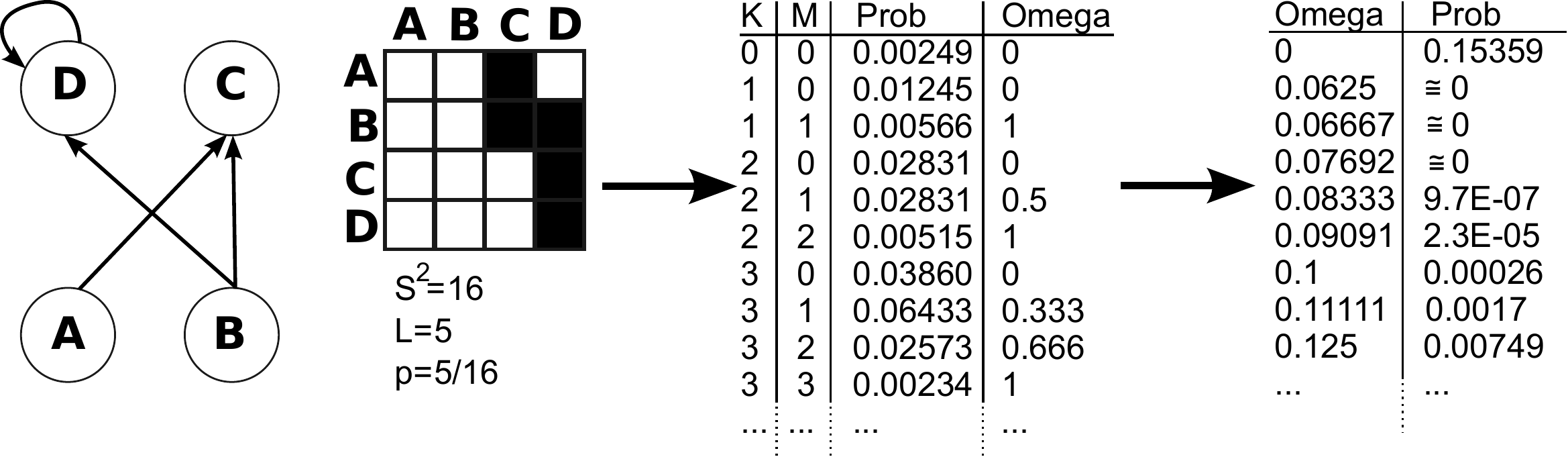}
  \end{center}
  \caption{Building the exact probability mass distribution for the
    overlap of links using a random digraph. First, evaluate relevant
    parameters (left). Then, build a table for all the possible
    combinations of $K$ and $M$ (center, just 10 of the 87 rows
    presented). Finally, condense the table according to $\Omega$,
    creating a univariate pmf (right).}
\end{figure}

\newpage

\begin{figure}
  \begin{center}
    \includegraphics[width=0.8\textwidth]{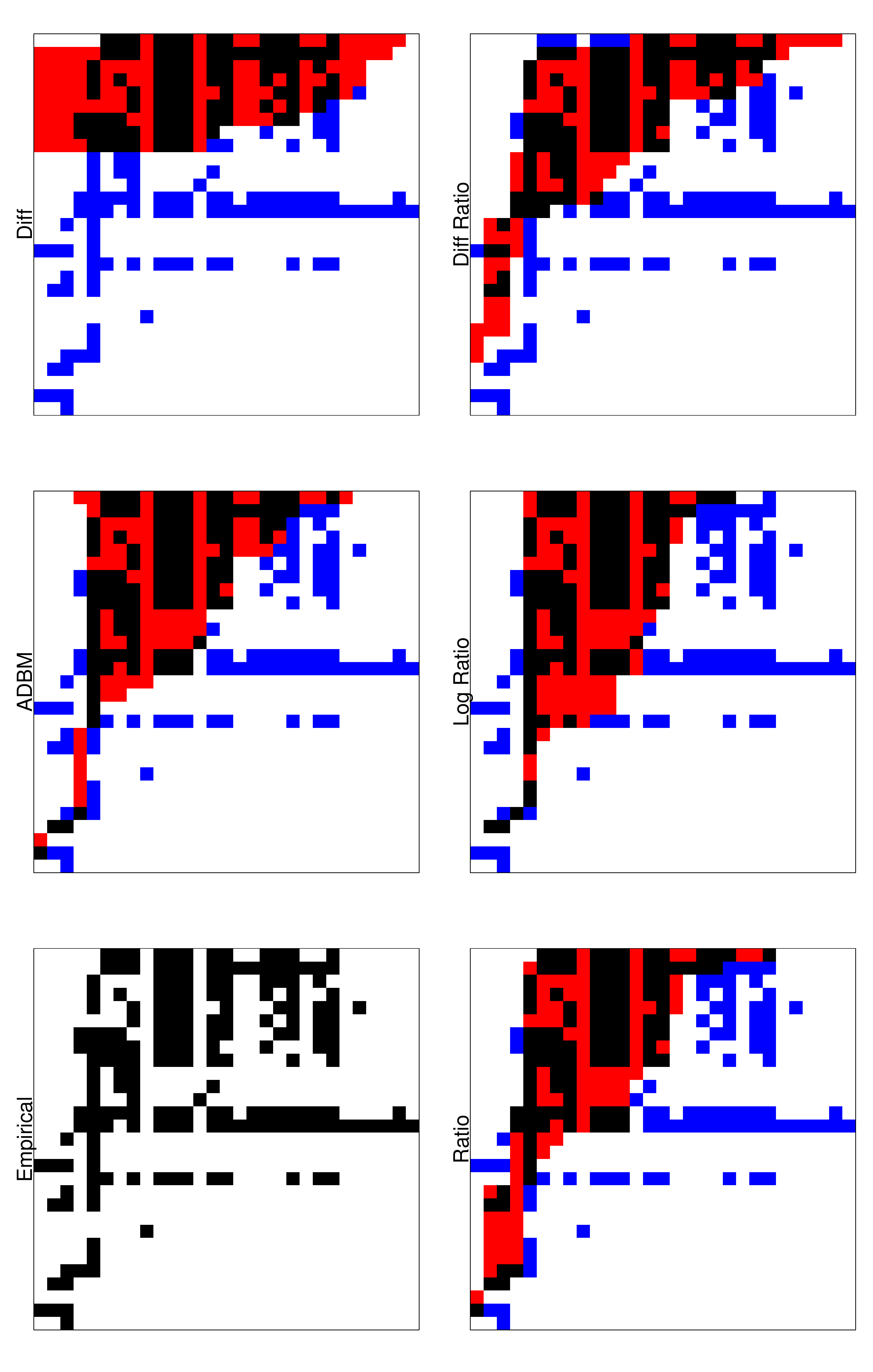}
  \end{center}
  \caption{Benguela Pelagic food web. For each model, I report the links
    correctly predicted (black), those incorrectly predicted (red) and
    those not predicted by the model but present in the empirical web
    (blue).}
\end{figure}

\newpage

\begin{figure}
  \begin{center}
    \includegraphics[width=0.8\textwidth]{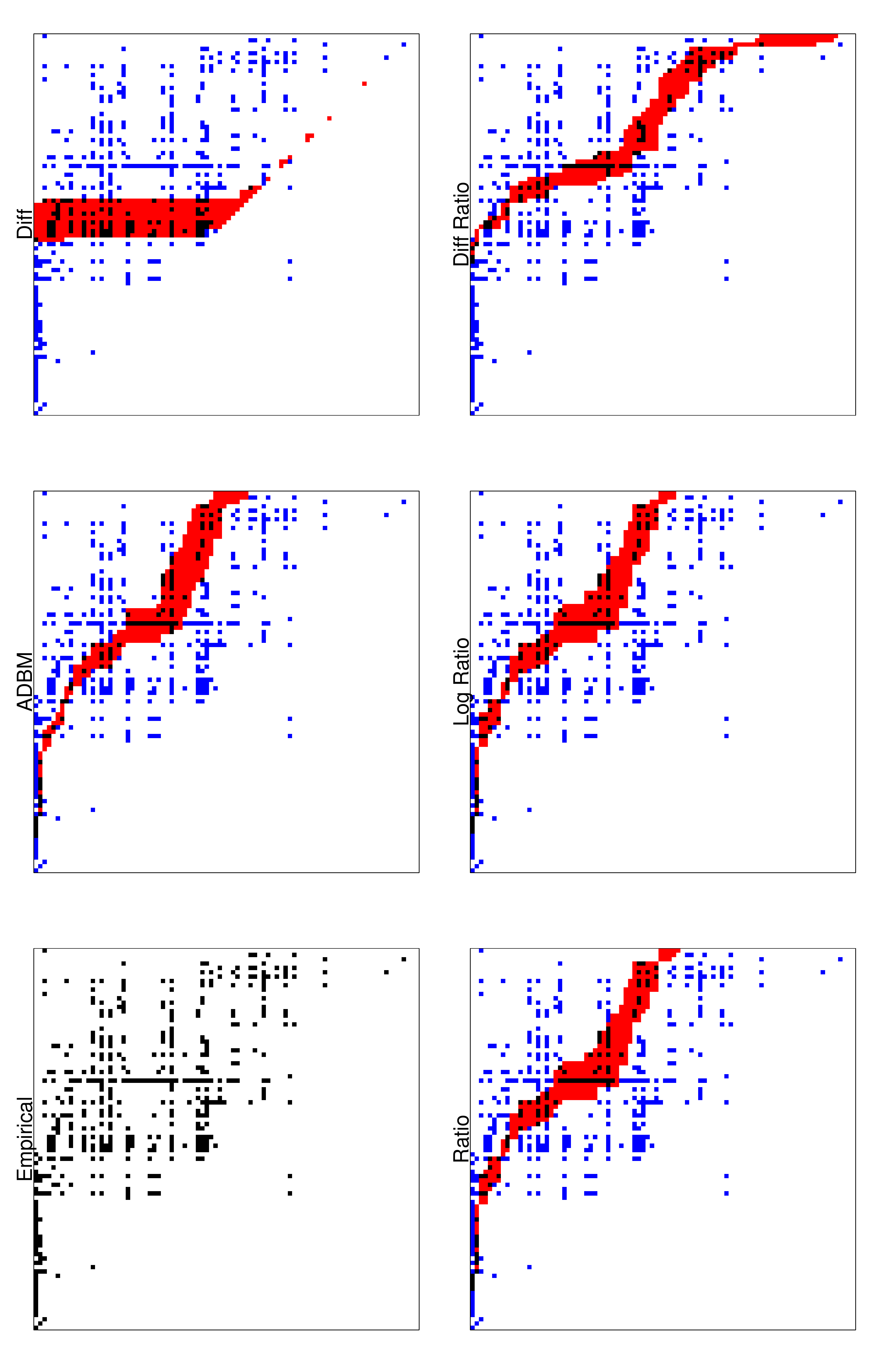}
  \end{center}
  \caption{Ythan Estuary food web. For each model, I report the links
    correctly predicted (black), those incorrectly predicted (red) and
    those not predicted by the model but present in the empirical web
    (blue).}
\end{figure}

\newpage

\begin{table}
\begin{tabular}{|l|l|l|c|c|c|c|c|c|}
\hline
Food Web	&$S$	&$L$	&$\underset{ADBM}{\Omega}$	
&$\underset{Diff}{\Omega}$	&$\underset{Ratio}{\Omega}$	&$\underset{LogRatio}{\Omega}$	
&$\underset{DiffRatio}{\Omega}$\\ \hline
Benguela	&29	&191	&0.57143	&0.48705	&0.56771	&0.557895	&0.54497	\\
Broadstone	&29	&156	&0.40385	&0.42308	&0.38461	&0.384615	&0.40385	\\
Broom	&68	&101	&0.07767	&0.1	&0.13592	&0.137255	&0.09804	\\ 
Carpinteria	&72	&238	&0.16456	&0.21429	&0.16318	&0.172996	&0.15900	\\ 
Coachella	&26	&228	&0.65065	&0.52863	&0.63877	&0.656388	&0.57205	\\ 
Sierra	&33	&175	&0.60366	&0.61047	&0.50610	&0.487805	&0.55758	\\ 
Skipwith	&71	&347	&0.13833	&0.12680	&0.13256	&0.132565	&0.13833	\\ 
Tuesday	&73	&410	&0.46472	&0.40146	&0.46472	&0.462287	&0.43796	\\ 
Ythan	&88	&425	&0.18824	&0.21177	&0.20235	&0.202353	&0.17412
\\ \hline
\end{tabular}
\label{Omegas}
\caption{Overlap values for the ADBM and the four simpler models based on body size described in the text.}
\end{table}

\newpage

\begin{table}
\begin{tabular}{|l|c|c|c|c|c|c|}
\hline
Food Web	&$\underset{ADBM'}{AIC}$	&$\underset{Diff'}{AIC}$	
&$\underset{Ratio'}{AIC}$	&$\underset{LogRatio'}{AIC}$	&$\underset{DiffRatio'}{AIC}$	&$\underset{ADBM'}{A.W.}$\\
\hline
Benguela	&825.25	&880.56	&821.08	&831.32	&842.5	&1.100E-01\\
Broadstone	&824.63	&811.5	&829.04	&829.04	&820.63	&1.400E-03\\
Broom	&1110.72	&1100.12	&1085.6	&1085.34	&1100.45	&1.640E-06\\
Carpinteria	&2036.7	&1991.09	&2033.25	&2026.37	&2036.29	&1.250E-10\\
Coachella	&777.31	&869.37	&786.82	&769.85	&840.33	&2.340E-02\\
Sierra	&823.69	&798.66	&900.53	&913.57	&858.9	&3.670E-06\\
Skipwith	&2658.12	&2660.53	&2657.44	&2657.44	&2654.12	&8.700E-02\\
Tuesday	&2516.69	&2654.99	&2512.69	&2518.53	&2575.28	&1.140E-01\\
Ythan	&3380.31	&3343.62	&3357.12	&3357.12	&3394.18	&1.070E-08\\ \hline
\end{tabular}
\caption{AIC values for the probabilistic extensions of the ADBM and the
  other four simpler models described in the text. The AIC accounts for
  the number of parameters as well as the goodness of fit. Akaike
  weights (A.W.) measure the confidence that the ADBM is the best among
  the examined models.}
\end{table}

\newpage

\begin{sidewaystable}
  \footnotesize
  \begin{center}
    \begin{tabular}{|l|c|c|c|c|c|c|c|} \hline
      Food Web	&$\underset{ADBM}{\Omega}$	
 &${\cal P}\left(\underset{RND}{\Omega}\geq \underset{ADBM}{\Omega} \right)$	& $E\left[{\underset{RND}{\Omega}}\right]$
 &${\cal P}\left(\underset{CASC}{\Omega}\geq \underset{ADBM}{\Omega} \right)$	& $E\left[{\underset{CASC}{\Omega}}\right]$
 &${\cal P}^*\left(\underset{GROUP}{\Omega}\geq \underset{ADBM}{\Omega} \right)$	
 & $E^*\left[{\underset{GROUP}{\Omega}}\right]$
      \\ \hline

      Benguela	&0.571	&2.289E-27	&0.227	&6.123E-09	&0.411	&1.000E+00	&0.727\\
      Broadstone &0.404	&3.965E-12	&0.185	&8.152E-04	&0.306	&1.000E+00	&0.858\\
      Broom	&0.078	&1.145E-03	&0.022	&6.038E-02	&0.044	&1.000E+00	&0.386\\
      Carpinteria &0.165	&1.714E-12	&0.046	&1.246E-04	&0.093	&1.000E+00	&0.396\\
      Coachella	& 0.651	&4.645E-28	&0.337	&2.070E-07	&0.531	&1.000E+00	&0.834\\
      Sierra	&0.604	&6.327E-37	&0.161	&3.992E-19	&0.317	&1.000E+00	&0.855\\
      Skipwith	 &0.138	&1.155E-06	&0.069	&4.515E-01	&0.136	&1.000E+00	&0.559\\
      Tuesday	 &0.465	&1.213E-83	&0.077	&1.045E-53	&0.149	&1.000E+00	&0.833\\
      Ythan	&0.188	&1.010E-22	&0.055	&1.474E-07	&0.109	&1.000E+00	&0.447\\
      \hline
    \end{tabular}
  \end{center}
  \normalsize
  \caption{Size ($S$) and number of connections ($L$) in nine empirical
    networks. For each of the probabilistic models, I report both the
    probability that they produce an overlap greater than that of the
    ADBM and their expected overlap. The last two values (marked with
    $*$) have been obtained through simulations, because the exact
    computation is not feasible.}
  \label{Tablepvalues}
\end{sidewaystable}

\newpage

\begin{table}
  \begin{tabular}{|l|c|c|c|c|c|c|c|} \hline
    Food Web	& $\log{\underset{RND}{\cal L}}$	& $\underset{RND}{AIC}$
    	& $\log{\underset{CASC}{\cal L}}$	& $\underset{CASC}{AIC}$
    	& $\log{\underset{GROUP}{\cal L}}$	& $\underset{GROUP}{AIC}$ & $\gamma$\\ \hline
Coachella	&-431.11	&866.22	&-350.4535	&704.907	&-141.637	&411.274	&8\\
Benguela	&-449.575	&903.15	&-364.77	&733.54	&-189.4495	&476.899	&7\\
Broadstone	&-402.36	&808.72	&-363.1855	&730.371	&-99.9415	&271.883	&6\\
Sierra	&-479.06	&962.12	&-388.266	&780.532	&-107.6645	&313.329	&7\\
Broom	&-485.1	&974.2	&-480.4365	&964.873	&-279.8005	&657.601	&7\\
Skipwith	&-1262.36	&2528.72	&-1097.235	&2198.47	&-559	&1360	&11\\
Carpinteria	&-964.745	&1933.49	&-862.555	&1729.11	&-536.475	&1272.95	&10\\
Tuesday	&-1444.36	&2892.72	&-1254.93	&2513.86	&-295.6795	&833.359	&11\\
Ythan	&-1645.715	&3295.43	&-1444.84	&2893.68	&-828.445	&1898.89	&11\\ \hline

  \end{tabular}
  \caption{ Likelihood and AIC values for all the networks using the
    three probabilistic models described in the main text.  The AIC
    takes into account the number of parameters that is 2 for the random
    digraph ($RND$), 2+$S$ for the cascade model ($CASC$) and
    2+$S$+$\gamma^2$ in the group-based random digraph. Because $\gamma$
    varies among networks, its value is reported as well.  }
  \label{TableAICValues}
\end{table}

\newpage
\section*{Appendix}

\subsection*{p-value: a cascade model}
Here I repeat the analysis above for a version of the cascade model. The
cascade model was the first probabilistic model for food web structure
to be proposed \citep{Cohen90FW}. I examine here a simple variation on
the original model. To produce a network, a vector $\vec{H}$
representing a hierarchy (an order) of the species is required. If we
order the empirical network according to $\vec{H}$, we can divide the
links in the network into two classes: a) connections from lower ranked
species to higher ranked species (forward connections) and b)
connections from higher to lower or equal ranked species (backward
connections). In the adjacency matrix associated with the ordered
network, the forward connections are contained in the upper triangular
part of the matrix, while the backward connections lie either on the
lower triangular part or on the diagonal. Having set the number of
species and the hierarchy among them, we connect species in the
following way: we draw forward connections with probability $p_U$ and
backward connections with probability $p_L$. We define $L_U$ as the
number of links in the upper triangular part of the empirical network,
$L_L$ as the number of links in the lower part, $K_U$ and $K_L$ as the
number of links proposed by the model in the upper and lower part and
$M_U$ and $M_L$ as the matched links. It is trivial, given the
derivation for the random graph, to write the probability mass function
for this case:

\begin{equation} 
  \begin{split}
    {\cal P}(M_U,M_L,K_U,K_L|S,p_U,p_L, \vec{H},N(S,L_U,L_L))=\\
    p_U^{K_U}(1-p_U)^{\left(\frac{S(S-1)}{2}-K_U\right)}\binom{L_U}{M_U}\binom{\frac{S(S-1)}{2}-L_U}{K_U-M_U}\\
    p_L^{K_L}(1-p_L)^{\left(\frac{S(S+1)}{2}-K_L\right)}\binom{L_L}{M_L}\binom{\frac{S(S+1)}{2}-L_L}{K_L-M_L}\\
\end{split}
\end{equation}

Where $K_U \in [0, \ldots, \frac{S(S-1)}{2}]$, for each $K_U$ $M_U \in
[0, \ldots, min(K_U,L_U)]$, while $K_L \in [0, \ldots,
\frac{S(S+1)}{2}]$ and $M_L \in [0, \ldots, min(K_L,L_L)]$. The total
number of combinations for the four values of interest therefore can be
quite large:

\begin{equation}
  \begin{split}
    \text{Num. cases}=\left(\left(\frac{S(S-1)}{2}+1\right)(L_U+1)-\frac{L_U(L_U+1)}{2}
    \right)\cdot \\
    \left(\left(\frac{S(S+1)}{2}+1\right)(L_L+1)-\frac{L_L(L_L+1)}{2}
    \right)
    \end{split}
\end{equation}

 For example, for the Ythan estuary food web we have $L_U=421$,
$L_L=4$, $S=88$ leading to more than $2.989 \cdot 10^{10}$ possible
combinations. Although the number of combinations is very high, it is
still possible to compute the univariate distribution for $\Omega$ in
the same exact way as for the random graph by condensing the
multivariate distribution.

Also for this model one can easily derive the likelihood by setting
$K_U=M_U=L_U$ and $K_L=M_L=L_L$:

\begin{equation}
\begin{split}
  {\cal L}(S,p_U,p_L,\vec{H}|N(S,L_U,L_L))=\\
  p_U^{L_U} p_L^{L_L} (1-p_U)^{\frac{S(S-1)}{2}-L_U}(1-p_L)^{\frac{S(S+1)}{2}-L_L}
\end{split}
\end{equation}

And the AIC:

\begin{equation}
\begin{split}
  AIC=6+2S-2\left(L_U\, log(p_U)+\left(\frac{S(S-1)}{2}-L_U\right)\,log(1-p_U)\right)\\
  -2\left(L_L\, log(p_L)+\left(\frac{S(S+1)}{2}-L_L\right)\,log(1-p_L)\right)
\end{split}
\end{equation}

\subsection*{p-value: a group-based random digraph}
Finally, I derive here the probability of obtaining any $\Omega$ for a
model that is a collection of random digraphs in which species interact
according to the ``group'' they belong to \citep{AllesinaGroups}. For
example, if we divide the nodes of a network into two groups (``red''
and ``green''), we will use four probabilities for deciding whether to
connect a red node to a red node ($p_{rr}$), a red node to a green node
($p_{rg}$), a green to a green ($p_{gg}$) and a green to a red
($p_{gr}$). The number of probabilities required will therefore be
$\gamma^2$ where $\gamma$ is the number of groups. This model is simply
a collection of random subgraphs. We first define a vector $\vec{G}$
containing, for each species, the group the species is assigned to. We
further define $L_{ij}$ as the number of links in the empirical network
connecting resources belonging to the $i^{th}$ group to consumers
belonging to the $j^{th}$ group, $K_{ij}$ as the number of links
proposed by the model for the interaction between these groups and
$M_{ij}$ the matched links. Finally, we write $<i>$ for the size of the
$i^{th}$ group. We can now write the multivariate pmf for all
combinations of $K_{ij}$ and $M_{ij}$:

\begin{equation}
  \begin{split}
    {\cal P}(\vec{K}_{ij},\vec{M}_{ij}|S,\vec{p}_{ij},\vec{G},N(S,\vec{L}_{ij}))=\\
    \prod_i^\gamma \prod_j^\gamma \left[
      p_{ij}^{K_{ij}}(1-p_{ij})^{<i><j>-K_{ij}} \binom{L_{ij}}{M_{ij}}
      \binom{<i><j>-L_{ij}}{K_{ij}-M_{ij}} \right]
  \end{split}
\end{equation}

Note that the model is conceptually very simple: in the case $\gamma=1$
the model reduces to the random digraph described above. Although
listing all the possible cases is theoretically feasible, their number
can be immense:

\begin{equation}
  \text{Num. cases}= \prod_i^\gamma \prod_j^\gamma \left[(<i><j>+1)(L_{ij}+1)-\left(\frac{L_{ij}(L_{ij}+1)}{2}\right) \right]
\end{equation}

For example, for the Coachella Valley food web examined below, I found
more than $10^{43}$ possible combinations, so that obtaining the exact
distribution is not computationally feasible. Nevertheless, as for the
other cases, the likelihood and the AIC are readily derived and easy to
compute:

\begin{equation}
  \begin{split}
    {\cal L}(S,\vec{p}_{ij},\vec{G}|N(S,\vec{L}_{ij}))=
    \prod_i^\gamma \prod_j^\gamma \left[
      p_{ij}^{L_{ij}}(1-p_{ij})^{<i><j>-L_{ij}} \right]
  \end{split}
\end{equation}

\begin{equation}
  \begin{split}
    AIC=2 +2S+2\gamma^2-2\sum_i^\gamma \sum_j^\gamma \left[
      L_{ij} log (p_{ij})+(<i><j>-L_{ij})log(1-p_{ij}) \right]
  \end{split}
\end{equation}

\end{document}